\documentclass[aps,pra,groupedaddress,twocolumn,amsmath,
amssymb, floatfix]{revtex4-1}
\usepackage{graphicx}
\usepackage{bm}
\usepackage{verbatim}
\usepackage{amsmath}
\usepackage{dcolumn}
\usepackage{xcolor}
\usepackage{SIunits}
\usepackage{booktabs}

\usepackage{tikz}
\definecolor{darkred}{rgb}{0.545, 0.0, 0.0}
\newcommand{\bluecircle}[0]{%
  \begin{tikzpicture}[x=1pt, y=1pt]
    \draw[fill=blue, thick] (0, -2) circle (2.5);
  \end{tikzpicture}%
}
\newcommand{\redsquare}[0]{%
  \begin{tikzpicture}[x=1pt, y=1pt]
    \draw[darkred, very thick] rectangle (4.5, 4.5);
  \end{tikzpicture}%
}

\newcommand{\greendiamond}[0]{%
  \begin{tikzpicture}[x=1pt, y=1pt]
    \draw[fill=green!50!black, thick, rotate=45] rectangle (4.5, 4.5);
  \end{tikzpicture}%
}

\newcommand{\greendiamondempty}[0]{%
  \begin{tikzpicture}[x=1pt, y=1pt]
    \draw[green!50!black, very thick, rotate=45] rectangle (4.5, 4.5);
  \end{tikzpicture}%
}

\newcommand{\ket}[1]{\vert#1\rangle}

\newcommand{\projector}[2]{\vert #1 \rangle\hspace{-0.3ex} \langle #2
  \vert}

\newcommand{\mycite}[1]{~\cite{#1}}
\newcommand{\myciten}[1]{~\cite{#1}}

\newcommand{\apdxname}{Appendix}

\begin{document}

\title{Heralded quantum entanglement between two crystals}

\author{Imam~Usmani} \thanks{These authors contributed equally to this
  work.}  \author{Christoph~Clausen} \thanks{These authors contributed
  equally to this work.}  \author{F\'{e}lix Bussi\`{e}res}
\thanks{These authors contributed equally to this work.}
\author{Nicolas~Sangouard} \author{Mikael~Afzelius}
\email{mikael.afzelius@unige.ch} \author{Nicolas Gisin}

\affiliation{Group of Applied Physics, University of Geneva, CH-1211
  Geneva 4, Switzerland}%

\date{24 August 2011}

\begin{abstract}
  Quantum networks require the crucial ability to entangle quantum
  nodes\mycite{Kimble2008}. A prominent example is the quantum
  repeater\mycite{Briegel1998,Duan2001,Sangouard2011} which allows
  overcoming the distance barrier of direct transmission of single
  photons, provided remote quantum memories can be entangled in a
  heralded fashion. Here we report the observation of heralded
  entanglement between two ensembles of rare-earth-ions doped into
  separate crystals. A heralded single photon is sent through a 50/50
  beamsplitter, creating a single-photon entangled state delocalized
  between two spatial modes. The quantum state of each mode is
  subsequently mapped onto a crystal, leading to an entangled state
  consisting of a single collective excitation delocalized between two
  crystals. This entanglement is revealed by mapping it back to
  optical modes and by estimating the concurrence of the retrieved
  light state\mycite{Chou2005}. Our results highlight the potential of
  rare-earth-ions doped crystals for entangled quantum nodes and bring
  quantum networks based on solid-state resources one step closer.
\end{abstract}

\maketitle

Quantum entanglement challenges our intuition about physical reality.
At the same time, it is an essential ingredient in quantum
communication\mycite{Gisin2007a}, quantum precision
measurements\mycite{Giovannetti2004} and quantum
computing\mycite{Nielsen2000,Kok2007}. In quantum communication, photons
are naturally used as carriers of entanglement using either free-space
or optical fiber transmission. Yet, even with ultra low-loss
telecommunication optical fiber, the transmission probability
decreases exponentially with distance, limiting the achievable
communication distance to a few hundred kilometers\mycite{Gisin2002}. A
potential solution is to use quantum repeaters\mycite{Briegel1998} based
on linear optics and quantum memories\mycite{Duan2001,Sangouard2011}
with which the entanglement distribution time scales polynomially with
the transmission distance, provided entanglement between quantum
memories in remote locations can be heralded. In this context, one can
more generally consider prospective quantum networks\mycite{Kimble2008}
where nodes generate, process and store quantum information, while
photons transport quantum states from site to site and distribute
entanglement over the entire network. An essential step towards the
implementation of such potential technologies is to create
entanglement between two quantum memories in an heralded
manner\mycite{Duan2001,Sangouard2011}.

Experimental observation of heralded entanglement between two
independent atomic systems for quantum networks was achieved using
cold gas ensembles involving either two distinct
ensembles\mycite{Chou2005}, or two spatial modes in the same
ensemble\mycite{Julsgaard2001,Simon2007c, Laurat2007, Choi2008,
  Matsukevich2006a} (note that in ref.\myciten{Matsukevich2006a} the
entanglement was postselected). Complete elementary links of a quantum
repeater (based on heralded entanglement) have been implemented in
cold gas systems\mycite{Chou2007, Yuan2008} and with two trapped
ions\mycite{Moehring2007}. For the realization of scalable quantum
repeaters, solid-state devices are technologically
appealing\mycite{Tittel2010}. In this context, important results have
already been obtained with rare-earth-ion (RE) doped crystals used as
quantum memories. These results include light storage times greater
than one second\mycite{Longdell2005}, storage efficiency of
69\%\mycite{Hedges2010} and quantum storage of~64 independent optical
modes in one crystal\mycite{Usmani2010} (see also
ref.\myciten{Bonarota2011}). Recent achievements\mycite{Clausen2011,
  Saglamyurek2011} include the storage, in a single RE-doped crystal,
of photonic time-bin entanglement generated through spontaneous
parametric downconversion (SPDC). Heralded entanglement between two
RE-doped crystals using an SPDC source, a common resource in quantum
optics, represents an important step towards the implementation of
quantum repeater architectures based on solid-state
devices\mycite{Simon2007}.

\begin{figure*}[!]
  \includegraphics[width=12cm]{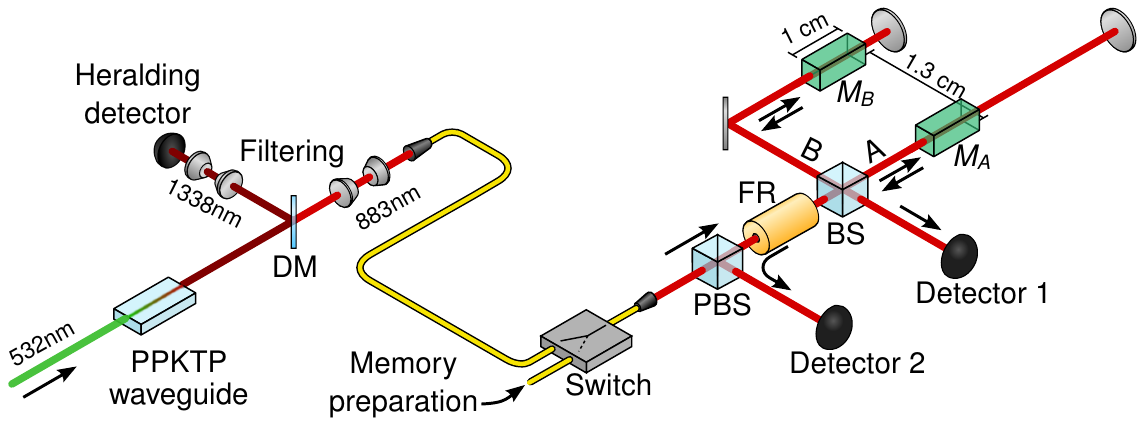}
  \caption{\textbf{Experimental setup.}  The quantum memories $M_A$
    and $M_B$ are implemented using neodymium ions doped into yttrium
    ortho-silicate crystals (Nd$^{3+}$:Y$_2$SiO$_5$) separated by
    1.3~cm and cooled to 3~K using a cryostat (see
    ref.\myciten{Clausen2011} for details). The total efficiency of
    each memory was 15\%. A fiber optic switch is used to alternate
    between a 15~ms long preparation of the two Nd ensembles as atomic
    frequency combs on the $^4$F$_{3/2} \rightarrow ^4$I$_{9/2}$
    transition, followed by attempts of entanglement creation for
    another 15~ms. For the latter, continuous wave light at 532~nm is
    coupled into a periodically poled KTP waveguide, leading to the
    production of pairs of photons at wavelengths of 883~nm and
    1338~nm through spontaneous parametric downconversion. Photons
    from each pair are separated on a dichroic mirror (DM) and
    frequency filtered to below the 120~MHz bandwidth of the quantum
    memories. The detection of an idler photon at 1338~nm (using a
    low-noise superconducting single photon
    detector\mycite{Verevkin2004}) heralds the presence of a signal
    photon at 883~nm. The signal photon now traverses the switch, a
    polarizing beam splitter (PBS) and a Faraday rotator (FR), before
    a 50/50 beamsplitter (BS) creates single-photon entanglement
    between spatial modes $A$ and $B$. This entanglement is, upon
    absorption, mapped onto the crystals $M_A$ and $M_B$. After a
    preprogrammed storage time of 33~ns, the photons are reemitted and
    pass through the BS again. Depending on which output mode of BS
    they emerge from, they either reach detector~1 or are rotated in
    polarization by the FR and reflected by the PBS towards detector
    2.}
  \label{fig:interferometer}
\end{figure*}

Here we present the observation of heralded quantum entanglement
between two neodymium ensembles doped in two yttrium ortho-silicate
crystals (Nd$^{3+}$:Y$_2$SiO$_5$) separated by 1.3~centimeters.  Let
us conceptually detail our experiment, depicted in Fig.~1. A
nonlinear optical waveguide is pumped to produce photon pairs through
SPDC. The resulting idler (1338~nm) and signal (883~nm) photons are
strongly filtered to match the absorption bandwidth of the crystals,
yielding a coherence time of 7~ns.  In the limit where the probability
of creating a single pair is much smaller than one, the detection of
an idler photon heralds the presence of a single signal photon. By
sending the latter through a balanced beamsplitter, one ideally
heralds a single-photon entangled state\mycite{Enk2005}
$\frac{1}{\sqrt{2}}(\ket{1}_A \ket{0}_B + \ket{0}_A\ket{1}_B)$ between
the two spatial output modes $A$ and $B$. In each of the modes a
crystal acts as quantum memory, denoted $M_A$ or $M_B$.  Upon
absorption of the single photon by one of the
crystals\mycite{Afzelius2009a}, the detection of an idler photon heralds
the creation of a single collective excitation delocalized between the
two crystals, ideally written as
$\frac{1}{\sqrt{2}}(\ket{W}_A\ket{0}_B+\ket{0}_A\ket{W}_B)$. Here,
$\ket{W}_A$ (or $\ket{W}_B$) is the Dicke-like state created by the
absorption of a single photon in $M_A$ (or $M_B$).  To determine the
presence of entanglement between the memories, we use a photon echo
technique based on an atomic frequency
comb\mycite{Afzelius2009a,Riedmatten2008,Afzelius2010,Sabooni2010,Bonarota2010}
that reconverts the collective excitation into optical modes $A$ and
$B$ after a preprogrammed storage time of 33~ns. Note that the
memories are used in a double-pass configuration, which effectively
increases the optical depth and the quantum memory efficiency up to
15\%. The resulting fields can then be probed using single photon
detectors to reveal heralded entanglement between the memories. Since
the entanglement cannot increase through local operations on the
optical modes $A$ and $B$, the entanglement of the retrieved light
fields provides a lower bound for the entanglement between the two
memories.

The photonic state retrieved from the memories is described by a
density matrix $\rho$, including loss and noise, expressed in the Fock
state basis. To reveal entanglement in this basis, one cannot resort
to violating a Bell inequality given solely inefficient and noisy
single photon detectors. Instead, as shown in ref.\myciten{Chou2005}
(see also ref.\myciten{Laurat2007,Choi2008}), a tomographic approach
based on single photon detectors can be used. Indeed, it is possible
to determine the presence of entanglement between the retrieved fields
from the knowledge of the heralded probabilities $p_{mn}$ of detecting
$m$ photons in mode $A$ and $n$ in mode $B$, where $m,n \in \{0,1\}$,
and from the coherence between these modes. More precisely, one can
obtain a lower bound on the concurrence $C$ of the detected fields, a
measure of entanglement ranging from~0 for a separable state to~1 for
a maximally entangled state, through
\begin{equation}
  \label{C}
  C \geq \max(0,V(p_{01}+p_{10})-2\sqrt{p_{00}p_{11}}).
\end{equation}
The term $V$ is the interference visibility obtained by recombining
optical modes $A$ and $B$ on a 50/50 beamsplitter: it is directly
proportional to the coherence between the retrieved fields in
modes~$A$ and~$B$. To obtain a large concurrence, that is, a large
amount of entanglement, one should maximize $V$ (the coherence) and
$p_{10} + p_{01}$ (the probability to detect the heralded photon), and
minimize $p_{00}$ and $p_{11}$ (the probabilities of detecting
separable states $\ket{0}_A\ket{0}_B$ and $\ket{1}_A\ket{1}_B$
stemming from a lost signal photon and from two signal photons,
respectively). Implicit to this method is the assumptions that i) the
creation of more than two pairs is negligible, and ii) the
off-diagonal elements of $\rho$ with different number of photons
vanish (this is valid since no local oscillator providing a phase
reference was used; see ref.\myciten{Chou2005}).

To estimate $V$, $p_{00}$, $p_{10}$, $p_{01}$ and $p_{11}$, we used
the setup of Fig.~1 in the following way.  First, the visibility $V$
was measured by allowing the re-emitted delocalized photon to
interfere with itself using a balanced Michelson interferometer whose
phase was actively stabilized (see Fig.~1). Then, we blocked spatial
mode~$B$ to estimate $p_{10}$ by summing the number of detections on
detectors~1 and~2, conditioned on a heralding signal. Probability
$p_{01}$ is estimated similarly by blocking mode~$A$. Note that the
probability of getting simultaneous counts on both detectors was
negligibly small in front of the probability of a single
detection. Then, $p_{00}$ is estimated through normalization of the
total probability, $p_{00} + p_{10} + p_{01} \approx 1$, which is
justified by the fact that $p_{11} \ll p_{10} + p_{01} \ll p_{00}$
(see measured values in the \apdxname). To estimate
$p_{11}$, we used two different methods, described below.

In the first method, we use a direct measurement of threefold
coincidences, i.e.~involving all three detectors. Details about this
measurement are given in the \apdxname. With a
pump power of 16~mW and a coincidence time window of 10~ns, we
obtained $C^{\text{(MLE)}} = 6.3(3.8) \times 10^{-5}$ using a maximum
likelihood estimation (MLE) of the threefold coincidence probability,
and $C^{\text{(CE)}}= 3.9(3.8) \times 10^{-5}$ using a more
conservative estimation (CE); see~\apdxname. Both estimations yield a
concurrence that is greater than~0 by at least one standard deviation,
and show that entanglement was indeed present between the two
crystals. This measurement required 166~hours, a period in which two
threefold coincidences were observed. The prohibitively long
integration time of this method prevented us from attempting it with
lower pump powers (i.e.~for lower probability of creating more than
one pair). Hence, to study how the concurrence changes with pump
power, we used a second method based on twofold coincidences, which we
now describe.

In the second method, $p_{11}$ is estimated using a supplementary
assumption (see the \apdxname~for details). This approach is motivated by the
results obtained in ref.\myciten{Laurat2007}. Specifically, we assume
that all the observed detections stem from a two-mode squeezed-state,
and thus the measured zero-time cross-correlation $\bar{g}_{s,i}$ can
be written as $\bar{g}_{s,i} =1+1/p$, where $p \ll 1$ and $p^2$ are
interpreted as the probabilities of creating one and two photon pairs,
respectively.  We then proceed as follows to estimate~$p_{11}$. We
first measure the zero-time cross-correlation $g_{s,i}^A$ between
detections in the idler mode and the signal mode~$A$ (with mode~$B$
blocked). Then we measure $g_{s,i}^B$ in the same way (with mode~$A$
blocked) and verify that $g_{s,i}^A \approx g_{s,i}^B$. We calculate
the average of $g_{s,i}^A$ and $g_{s,i}^B$, denoted $\bar{g}_{s,i}$,
and estimate $p_{11}$ using
\begin{equation}
  \label{p11}
  p_{11}= \frac{4p_{10}p_{01}}{\bar{g}_{s,i}-1}.
\end{equation}
In the \apdxname, we provide justifications and additional measurements that
support our assumption and give evidence that it yields a lower bound
on the concurrence. In particular, we measured the second-order
autocorrelation of the signal (or idler) mode without storage to be
$g^{(2)}_{s,s} (0) = 1.81(2)$ (or $g^{(2)}_{i,i} (0) = 1.86(9)$),
which is very close to the ideal value of~2 associated with the
thermal photon statistics of a two-mode squeezed state. We also
measured the zero-time second-order autocorrelation function of the
heralded signal photon just before storage and obtained
\mbox{$g_{s,s|i}^{(2)}(0) = 0.061(4)$} for a pump power of 8~mW, which
is consistent with $p\ll 1$.  Note that this method of estimating
$p_{11}$ is based on twofold coincidences as opposed to threefold
coincidences for the other method. For our specific setup, this
resulted in a reduction of the measurement time by a factor of $10^6$
for similar statistical confidence on the concurrence. This method
requires no physical modifications to the optical circuit to measure
the different components of the retrieved fields, which simplifies its
implementation.

\begin{figure}
  \includegraphics[width=8cm]{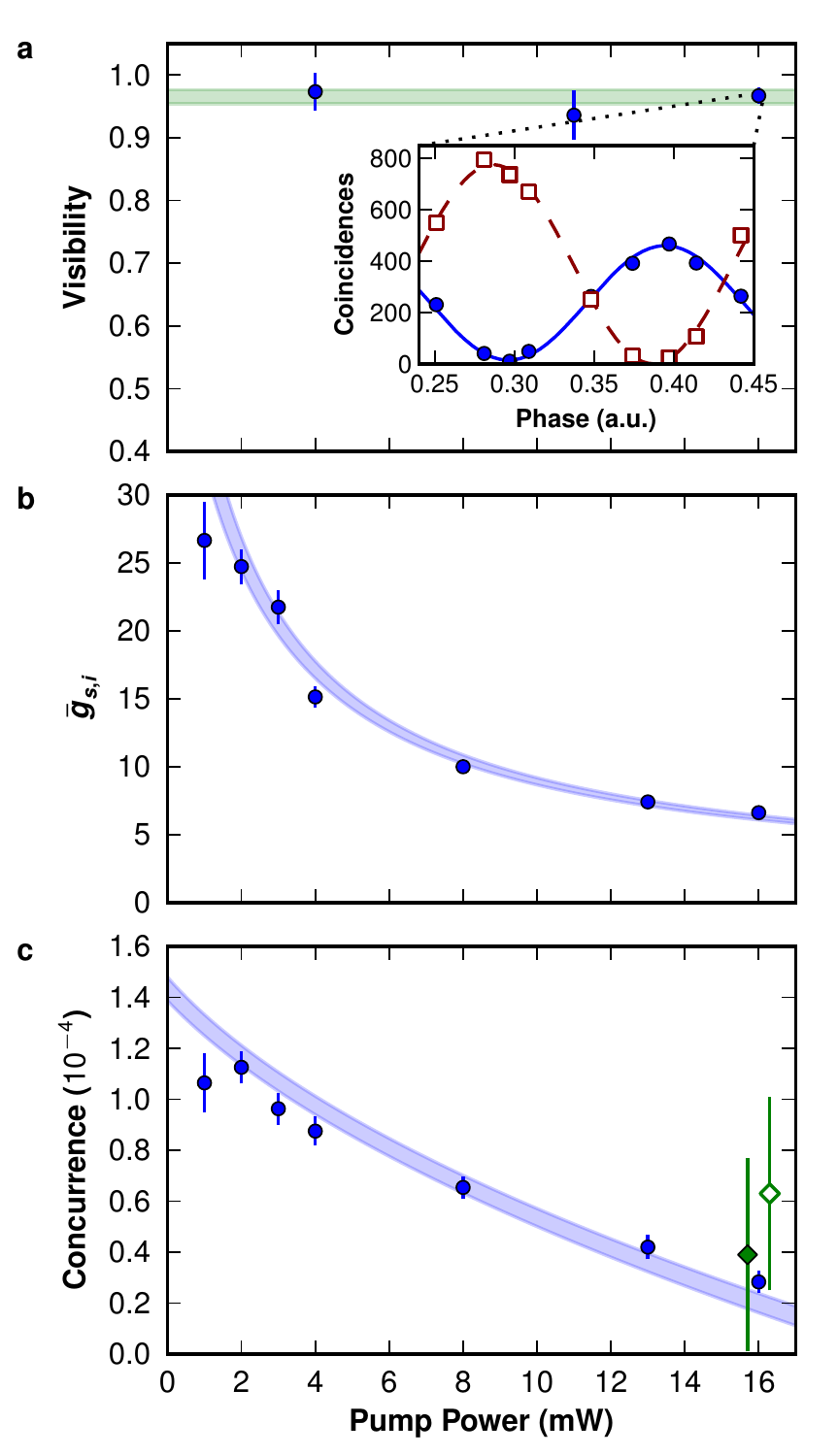}
  \caption[]{\textbf{Results.} \textbf{a},~Visibility as a function of
    pump power. The visibility is approximately constant with an
    average of $96.5\pm 1.2\%$ (green shaded region).  The inset shows
    the visibility curves at 16~mW measured with detectors~1
    (\bluecircle) and~2 (\redsquare) (15 minutes acquisition time per
    point; the error bars are smaller than the symbols). The different
    amplitudes result from non-uniform loss after recombination of
    modes $A$ and $B$ on the beamsplitter. The visibilities agree
    within uncertainties in the fits.  \textbf{b},~Zero-time
    cross-correlation $\bar{g}_{s,i}$ as a function of pump power. The
    decreasing values agree well with a theoretical model (shaded
    region; see \apdxname).  \textbf{c},~Lower bound on the concurrence
    estimated using the cross-correlation measurement the as a
    function of pump power (\bluecircle), calculated using Eq.~\ref{C}
    and~\ref{p11}, and the mean visibility of \textbf{a}. The
    concurrence decreases with pump power, as expected, but remains
    positive up to 16~mW.  The shaded region corresponds to the model
    in \textbf{b}. All measured values of $p_{10}$, $p_{01}$ and
    $p_{11}$ are given in the \apdxname.  All values are based on raw
    counts. Uncertainties are obtained assuming Poissonnian detection
    statistics.  The lower bound on the concurrence at 16~mW obtained
    from measured threefold coincidences using either the maximum
    likelihood estimation (\greendiamondempty) or the
    conservative estimation (\greendiamond) are also shown (they are
    horizontally offsetted for clarity).}
  \label{fig:results}
\end{figure}

We performed a series of measurements for several values of the pump
power, which is proportional to $p$ provided $p \ll 1$. For all
measurements we used coincidence windows of 10~ns, and all results are
based on raw counts (i.e.~without subtraction of dark counts and
accidental coincidences). The inset of 2a shows the interference of
the delocalized single photon retrieved from the memories as a
function of the phase of the interferometer. The visibilities obtained
with detectors~1 and~2 agree within the uncertainty of the fit. As
shown on Fig.~2a, the visibility does not depend on the pump power
and has an average of $96.5(1.2)\%$, implying almost perfect coherence
between the two memories and a large signal-to-noise ratio.  Fig.~2b
shows the measured $\bar{g}_{s,i}$. Reducing the pump power increases
the cross-correlation, as expected.  Using additional measurements, we
estimated the transmission loss, memory efficiency, dark count
probability and the pair creation probability of our setup. These
values were then used in a theoretical model (shaded region on
Fig.~2b) that is in excellent agreement with the measured values
of~$\bar{g}_{s,i}$. As explained in the \apdxname, this agreement provides
evidence that the estimated value of $p_{11}$ yields a lower bound on
the concurrence. Finally, Fig.~2c shows the lower bound on the
concurrence for all pump powers, calculated using Eqs.~\ref{C}
and~\ref{p11} and the average visibility of Fig.~2a. The concurrence
decreases with pump power because $p_{11}$ increases, while all other
terms of Eq.~\ref{C} depend on photon loss only and hence are
approximately constant. Nevertheless, the concurrence stays positive
for all used pump powers, proving the creation of heralded
entanglement between the atomic ensembles inside the two crystals. The
theoretical curve on Fig.~2c (shaded region) is based on the same
model used for the theoretical curve in Fig.~2b.

The results of the measurement of the concurrence based on threefold
coincidences are plotted on \mbox{Fig.~2c} and agree, within
uncertainty, with the result of the method based on measurement of the
cross-correlation.

The observed values of the concurrence are lower bounds on the amount
of entanglement of the detected fields and are almost entirely
determined by optical loss. To see this, we first note that when both
the multi-pair creation probability and the total transmission
probability $\eta$ of the signal photon are small, the concurrence is
approximately given by $C \approx \eta
(V-2/\sqrt{\bar{g}_{s,i}-1})$. At 8~mW of pump power, we measured
$g_{s,i} \approx 10$, $p_{00} = 0.9997831(71)$, and $p_{11} = 5.18(40)
\times 10^{-9}$, and hence $C\approx 0.3\eta$. With $\eta \approx
p_{10} + p_{01} = 2.2\times 10^{-4}$, we see that $C\approx 6.6\times
10^{-4} \ll 0.3$. Using this, an estimate of the concurrence of the
fields retrieved just after the crystals is obtained by noticing that
the transmission $\eta$ is the product of the probability to find the
signal photon inside the fiber per heralding signal (20\%), the memory
efficiency (15\%), the transmission in the interferometer (2.4\%) and
detector efficiency (30\%). Subtracting detector inefficiency and
interferometer loss yields a lower bound of approximately~$0.092$ for
the concurrence of the field retrieved just after crystals.

The high interference visibility indicates that the stored
entanglement does not significantly decohere during the 33~ns, and the
coherent nature of the storage endures for longer times, as shown
previously using storage of weak coherent
states\mycite{Riedmatten2008,Usmani2010}. Increasing the storage time
does however lower the cross-correlation function $\bar{g}_{s,i}$, as
measured in ref.\myciten{Clausen2011}, and this should consequently
lower the concurrence of the stored entanglement (this is essentially
due to the decrease of the memory efficiency with increasing storage
times, as shown in the \apdxname~and in ref.\myciten{Clausen2011}). With
Nd$^{3+}$:Y$_2$SiO$_5$, one could in principle store entanglement up
to $\sim 1$~\micro s (see ref.\myciten{Usmani2010}). A promising
approach to go beyond these limits, and to allow on-demand retrieval
of the stored entanglement, is to implement spin-wave
storage\mycite{Afzelius2009a} to increase storage times (as demonstrated
recently\mycite{Afzelius2010} in Pr$^{3+}$:Y$_2$SiO$_5$), and to place
the crystal inside an impedance-matched cavity to increase the
efficiency\mycite{Afzelius2010a}. Such improvements are necessary for
the development of a quantum repeater based on solid-state quantum
memories, and are the subject of current \mbox{research}.

In conclusion, we have reported an experimental observation of
heralded quantum entanglement between two separate solid-state quantum
memories. We emphasize that although the entangled state involves only
one excitation, the observed entanglement shows that the stored
excitation is coherently delocalized among all the neodymium ions in
resonance with the photon, meaning roughly $10^{10}$ ions in each
crystal. Our results demonstrate that rare-earth-ion ensembles,
naturally trapped in crystals, have the potential of compact, stable
and coherent quantum network nodes. Moreover, single photon
entanglement is a simple form of entanglement which can be used for
teleportation\mycite{Pegg1998} and entanglement swapping
operations\mycite{Duan2001} and can also be purified using linear
optical elements\mycite{Salart2010}. It is also a critical resource in
several proposals for quantum
repeaters\mycite{Duan2001,Simon2007,Sangouard2007a} since it is less
sensitive to transmission loss and detector
inefficiencies\mycite{Sangouard2011} as compared to two-photon entangled
states.

Our experimental approach is based on solid-state devices, the key
components being the PPKTP chip with an integrated waveguide (the
photon source) and the crystal memory. We believe this approach opens
possibilities of integration of components, for instance frequency
filtering directly on the chip\mycite{Pomarico2009}, or waveguide
quantum memories\mycite{Saglamyurek2011}. The prospect of combining
solid-state photon sources and quantum memories is attractive for
practical future quantum networks. One important challenge in this
context is to create entanglement between two remote solids in a
heralded way, which would realize an elementary link for quantum
repeaters.

\begin{acknowledgements}
  We thank H.~de Riedmatten, P.~Sekatski, and J.~Laurat for
  stimulating discussions, and A.~Korneev for help with the
  superconducting detector. This work was supported by the Swiss NCCR
  Quantum Science Technology, the Science and Technology Cooperation
  Program Switzerland-Russia, as well as by the European projects
  QuRep and ERC-Qore. F.B.~was supported in part by le Fond
  Qu\'eb\'ecois de la Recherche sur la Nature et les Technologies.
\end{acknowledgements}

\renewcommand{\thefigure}{A\arabic{figure}}
\renewcommand{\theequation}{A\arabic{equation}}
\renewcommand{\thetable}{A\arabic{table}}

\newcommand{\adag}[0]{a^{\dag}}
\newcommand{\bdag}[0]{b^{\dag}}
\newcommand{\ldag}[0]{l^{\dag}}

\appendix

\section*{Appendix}
This appendix provides details on the method we used to estimate the
threefold coincidence probability $p_{11}$ using either the
cross-correlation $g_{s,i}$ or by direct measurement. Specifically, in
section~\ref{section:p11gsi}, we present a model describing our setup
composed of the source of photon pairs and the memories, and includes
the effect of dark counts. We also detail additional measurements that
support our model. We present evidence that the method based on the
measurement of $g_{s,i}$ to estimate $p_{11}$ yields a lower bound on
the concurrence at detection and, consequently, on the concurrence of
the delocalized excitation stored in the quantum memories. In
section~\ref{section:p11direct}, we provide details on our method to
estimate $p_{11}$ directly, we show how the uncertainty on the
threefold coincidence probability is calculated and we show the
results.

\section{Concurrence using the
  cross-correlation} \label{section:p11gsi} 
\begin{figure}[t!]
  \includegraphics[width=\linewidth]{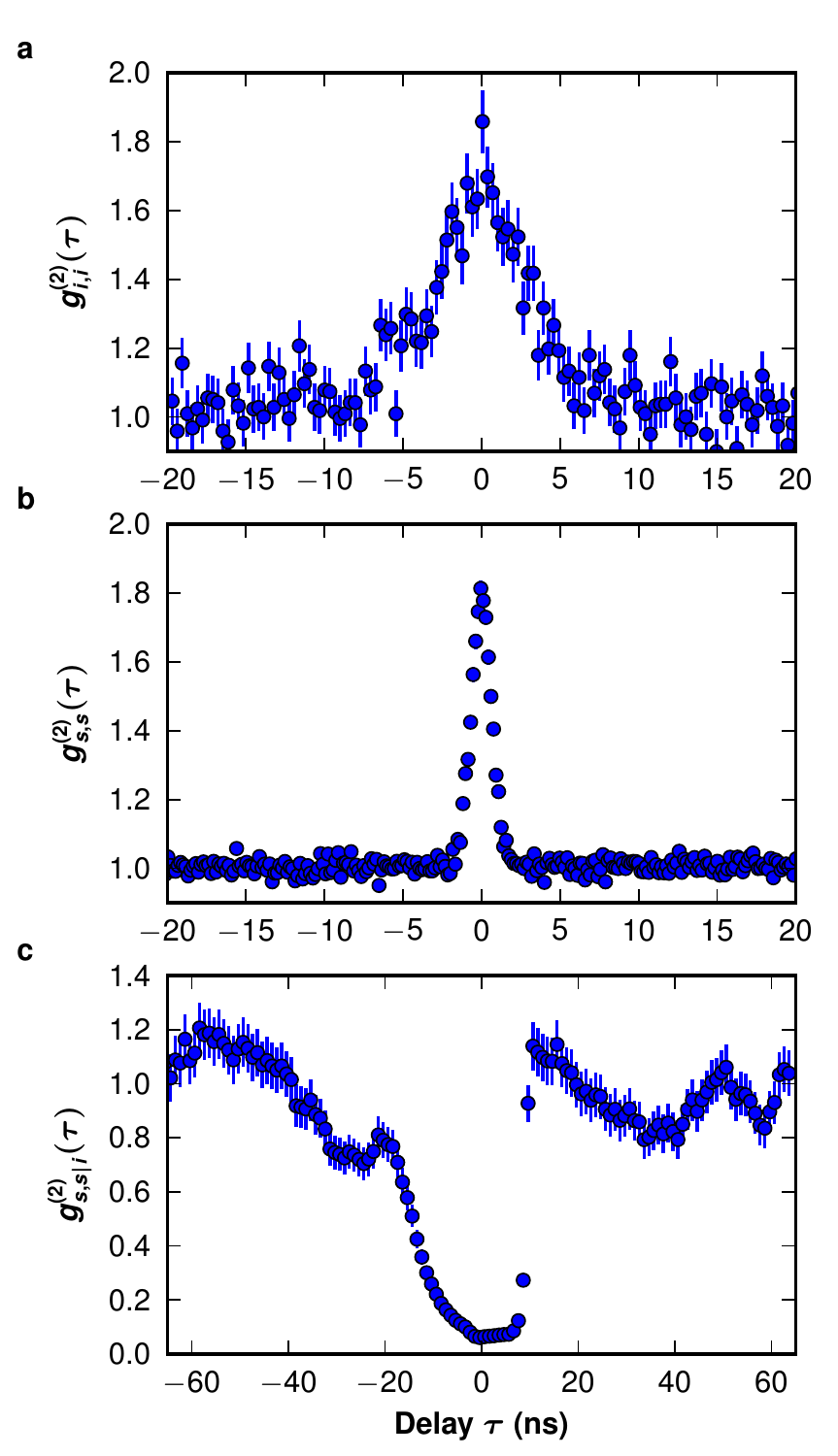}
  \caption{Measurements of second-order auto-correlation bunctions of
    \textbf{a}, the idler photon (1338~nm), \textbf{b}, the signal
    photon (883~nm) before it is sent towards the crystals, and
    \textbf{c}, the same as \textbf{b} but conditioned on a
    coincidence detection of an idler and the first signal photon at
    delay $\tau=0$.  The values at zero delay in \textbf{a} and
    \textbf{b} are close to the theoretical value of 2. The almost
    zero minimum value in \textbf{c} gives evidence of the single
    photon character of the heralded signal photon. Data was
    accumulated for 72 hours at a pump power of 16~mW in \textbf{a},
    and for 12 hours at 8~mW in \textbf{b} and \textbf{c}. The values
    in \textbf{c} have been calculated for a coincidence window of
    10~ns. Error bars are due to Poisson counting statistics.}
  \label{fig:g2source}
\end{figure}

Let us begin by a
description of the source. In our model, the SPDC source produces a
two-mode squeezed state (where the coherences are lost due to the lack
of a phase reference)
\begin{equation}
  \label{state_source}
  \cosh^{-2}{r} \sum_{n=0}^{+\infty} \tanh^{2n}{r} \-\ |n_in_s\rangle \langle n_in_s|.
\end{equation}
Here, $n_i$ and $n_s$ correspond to the number of photons in the idler
and signal modes (and are both equal to $n$).  While the signal and
idler fields are perfectly correlated in photon number, they
individually exhibit thermal statistics and hence, their second-order
auto-correlation function $g^{(2)}(0)$ is, in theory, equal to~2.  We
experimentally tested this for the photons emitted by the source by
performing direct measurements of $g_{i,i}^{(2)}(\tau)$,
$g_{s,s}^{(2)}(\tau)$ and $g_{s,s\vert i}^{(2)}(\tau)$, see
Fig.~\ref{fig:g2source}. The latter is the auto-correlation function
of the signal conditioned on the detection of an idler photon at
time~$\tau$. For the measurements we added 50/50 beam splitters after
the frequency filters of the signal or idler photon, respectively, and
registered coincidence events using free running detectors. We found
zero-delay auto-correlations of $g_{i,i}^{(2)}(0) = 1.86(9)$ and
$g_{s,s}^{(2)}(0) = 1.81(2)$, which are very close to the expected
values. This is consistent with the assumption that our source
produces a state of the form of
Eq.~\eqref{state_source}. Additionally, we found a conditional
auto-correlation of the signal photon at zero time delay of
$g_{s,s\vert i}^{(2)}(0) = 0.061(4)$ at a pump power of 8~mW. This
shows the very good single photon character of the heralded signal
fields, and that $\tanh^2 r \ll 1$.

\begin{figure}[t!]
  \centering
  \includegraphics{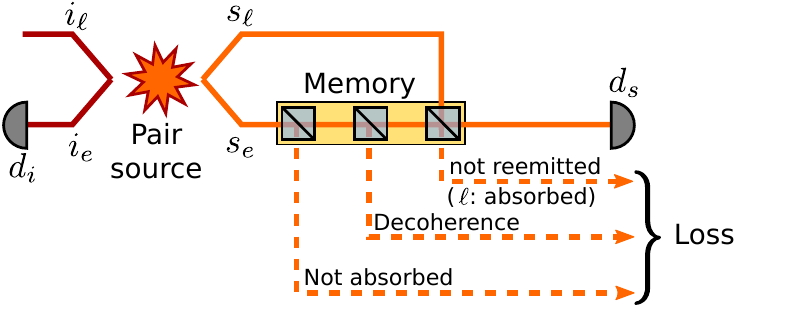}
  \caption{Model of our experimental setup. The source produces
    idler-signal pairs described by a two-mode squeezed state. The
    memory is represented by three beamsplitters (to account for the
    absorption, the decoherence and the re-emission
    respectively). Each temporal mode after the memory $d_s$ comes
    either from a signal photon emitted at an early time, then stored
    and later retrieved (mode $s_e$), or from a photon produced at a
    late time and directly transmitted through the memory (mode
    $s_\ell$). To take both contributions into account, a virtual
    source creating idler-signal pairs in a late time bin has been
    introduced. The correlations between the modes $d_s$ and $d_i$ are
    characterized through the measurement of the second-order
    cross-correlation function.}
  \label{fig:suppinfo}
\end{figure}
Let us continue the description of our model by focusing on the
memory. Due to non-unit absorption efficiency, each signal photon
detection after the memory stems either from a photon created at an
early time, stored in the medium, and later retrieved; or from a
photon created at a later time and not absorbed by the medium. Hence,
we need to consider two time bins. The storage process is a linear
operation, mapping the light field onto an atomic state. Thus, in our
model, it can be represented by a beamsplitter with a transmission
corresponding to the absorption probability,
c.f.~Fig.~\ref{fig:suppinfo}. Another beamsplitter is introduced to
account for the decoherence, inherent to the storage
process~\cite{Afzelius2009a}. Finally, a last beamsplitter accounts
for the reemission of the photon. It is identical to the first one and
combines the two time bins into a single temporal mode $d_s$. The
output of the last beamsplitter is detected to access the
cross-correlation function
\begin{equation}
  \bar g_{s,i} = \frac{\langle d_i^\dagger d_i d_s^{\dagger} d_s \rangle}{\langle d_i^\dagger d_i \rangle\langle d_s^\dagger d_s \rangle}
\end{equation}
between detection on the early idler mode $d_i$ and signal mode
$d_s$. Note that the overall transmission (in intensity) of the three
beamsplitters corresponds to the efficiency of the storage and
retrieval processes $\eta_{\text{echo}}$, while the reflection of the
last beamsplitter is the transmission of the atomic ensemble
$\eta_{\text{trans}}.$

Using the Heisenberg picture
$$
\langle d_s^\dagger d_s \rangle=\langle 0 | U^{\dagger} d_s^\dagger U
U^{\dagger} d_s U |0\rangle
$$
where
\begin{eqnarray}
  \nonumber
  &U^{\dagger}d_s U &=\sqrt{\eta_{\text{echo}}} \left(\cosh{r} \-\ s_e+ \sinh{r} \-\ i_e^\dagger\right)\\
  \nonumber
  &&+ \sqrt{\eta_{\text{trans}}} \left( \cosh{r} \-\ s_\ell+\sinh{r} \-\ i_{\ell}^\dagger \right)
\end{eqnarray}
leading to
$$
\langle d_s^\dagger d_s
\rangle=(\eta_{\text{trans}}+\eta_{\text{echo}}) \sinh^2{r}.
$$
Furthermore, we have
$$
U^{\dagger} d_i U = \left( \cosh{r} \-\ i_e+ \sinh{r} \-\
  s_e^\dagger\right)
$$
leading to
$$
\langle d_i^\dagger d_i \rangle= \sinh^2{r}.
$$
Following similar lines, we find
\begin{equation}
  \nonumber
  \begin{split}
    \langle d_i^\dagger d_i d_s^{\dagger} d_s \rangle=\sinh^2{r} \big(&\sinh^2{r} \left(\eta_{\text{trans}} +\eta_{\text{echo}}\right) \\
    &+\cosh^2{r} \-\ \eta_{\text{echo}}\big).
  \end{split}
\end{equation}
Therefore, the cross-correlation function between the idler-signal
modes is given by
$$
1+\frac{1}{\tanh^2{r}
  \left(1+\frac{\eta_{\text{trans}}}{\eta_{\text{echo}}}\right)}.
$$\\
Note that the effect of the memories is to add the term $(1 +
\frac{\eta_{\text{trans}}}{\eta_{\text{echo}}})$, which decreases the
cross-correlation.  One can further take the detector noise into
account by adding an additional source of noise with a Poissonian
photon distribution. We finally find
\begin{equation}
  \label{gsi_th}
  g_{s,i}=1+\frac{1}{\tanh^2{r} \left(1+\frac{\eta_{\text{trans}}}{\eta_{\text{echo}}}\right)+\frac{\eta_{\text{dark}}}{p_c}}
\end{equation}
where $\eta_{\text{dark}}$ is the probability to get a dark count
within the detection window and $p_c$ is the conditional probability
of detecting the signal photon retrieved from the memory. Note that
for small pump powers, the probability to create one pair is given by
\begin{equation*}
  \tanh^2 r/\cosh^2 r \approx r^2 = \alpha P_\text{pump},
\end{equation*}
where $P_\text{pump}$ is the continuous wave pump power.

To support the model presented above, we performed additional
measurements to determine the values of $\alpha$,
$\eta_{\text{trans}}/\eta_{\text{echo}}$, $p_c$ and (see
Table~\ref{tab:results}). First, we estimated $\alpha = 2.71(8)\times
10^{-3}~\text{pairs/mW}$ from a measurement of the cross-correlation
as function of pump power with one crystal prepared as a transmission
window (no storage) while the path of other crystal was blocked. Then,
$\eta_{\text{trans}}/\eta_{\text{echo}}$ was measured from the ratio
of the probabilities of detecting the heralded signal photon when it
is either transmitted through the memory (no storage), or stored and
retrieved from the memory. This measurement was performed twice --
once for each arm of the interferometer (with the other arm blocked)
-- and the outcomes were averaged. The conditional probability $p_c$
was taken as the average of probabilities $p_{10}$ and $p_{01}$ of
detecting the heralded signal photon retrieved from the memories (see
main text). Finally, we measured a dark count probability of
$\eta_{\text{dark}} = 2 \times 10^{-6}$ per 10~ns detection window for
the signal mode detector. Idler photons were detected using a
superconducting nanowire single photon detector with a dark count rate
of 20~Hz. This low dark count rate is neglected in our calculations.
Using these parameters, we compared the measured cross-correlation
$\bar{g}_{s,i}$ with values of $g_{s,i}$ predicted by
Eq.~\eqref{gsi_th}.  The
comparison 
is shown on Fig.~2B in the main text. We find excellent agreement
within the statistical uncertainties, thereby supporting our model for
the source and the memory.

\begin{table}
  \caption{Conditional probabilities $p_{01}$, $p_{10}$, $p_{11}$ 
    and the ratio $\eta_\text{trans} /
    \eta_\text{echo}$ for various pump powers.}
  \label{tab:results}
  \begin{tabular}{*{4}{m{.18\linewidth}}m{.18\linewidth}@{}}
    \toprule[1pt]
    \addlinespace[1pt]
    Pump power & 
    $p_{01}\;(10^{-4})$ &
    $p_{10}\;(10^{-4})$ &
    \multicolumn{1}{c}{$p_{11}\;(10^{-9})$} &
    \multicolumn{1}{c}{$\frac{\eta_\text{trans}}{\eta_\text{echo}}$} \\ 
    \addlinespace[1pt]
    \midrule[0.5pt]
    1 mW &   1.04(14) &  0.82(12) &  1.33(30) &  2.84(33) \\
    2 mW &  1.193(75) & 0.809(63) &  1.63(19) &  3.03(17) \\
    3 mW &  0.952(72) & 0.878(70) &  1.61(20) &  2.59(17) \\
    4 mW &  1.105(72) & 0.902(66) &  2.82(31) &  3.35(19) \\
    8 mW &  1.185(51) & 0.984(50) &  5.18(40) &  3.13(12) \\
    13 mW &  1.247(56) & 1.131(52) &  8.79(66) &  2.86(11) \\
    16 mW &  1.146(47) & 1.175(48) &  9.56(64) & 2.748(93) \\ 
    \midrule[0.5pt]
    \addlinespace[1pt]
    Mean & 1.123(30) & 0.957(27) &           & 2.936(69) \\ 
    \addlinespace[1pt]
    \bottomrule[1pt]
  \end{tabular}
\end{table}

We now give evidence based on our model that the method we used to
estimate the threefold coincidence probability leads to an
underestimation of the concurrence. For this, we compare the threefold
coincidence probability $p_{11}^{th}$ predicted by our model with the
one obtained from the method based on the measured cross-correlation
$\bar{g}_{s,i}$. Let us first concentrate on $p_{11}^{th}$. Using our
model, we can predict that the threefold coincidence probability
should be given by
$$
p_{11}^{th} = 4p_{10}p_{01} \left[ \alpha P_{\text{pump}} \left(
    1+\frac{\eta_{\text{trans}}}{2\eta_{\text{echo}}}\right)+\frac{\eta_{\text{dark}}}{p_c}.\right].
$$
This should be compared with the threefold coincidence probability
estimated using $\bar{g}_{s,i} = 1 + 1/p$, i.e.~assuming detections
stem from a two-mode squeezed states (see main text). Combining this
method with our model (Eq.~\ref{gsi_th}) yields
\begin{eqnarray*}
  p_{11} &=& \frac{4p_{10}p_{01}}{\bar g_{s,i}-1} \\
  &=& 4p_{10}p_{01} \left[ \alpha P_{\text{pump}} \left(1+\frac{\eta_{\text{trans}}}{\eta_{\text{echo}}}\right)+\frac{\eta_{\text{dark}}}{p_c}\right].
\end{eqnarray*}
One can see that the ratio $\eta_{\text{trans}} / \eta_{\text{echo}}$
enters in $p_{11}$ without the prefactor of one-half, such that
$p_{11} >p_{11}^{th}$. From Eq.~(1) of the main text, we directly see
that an overestimation of $p_{11}$ leads to a lower bound on the
concurrence at detection, and hence on the concurrence of the
entanglement stored in the memories.

\section{Concurrence using a direct measurement of the threefold
  coincidence probability} \label{section:p11direct}

\subsection{Method} \label{subsection:p11direct} 
\begin{figure*}[!t]
  \includegraphics[scale=0.85]{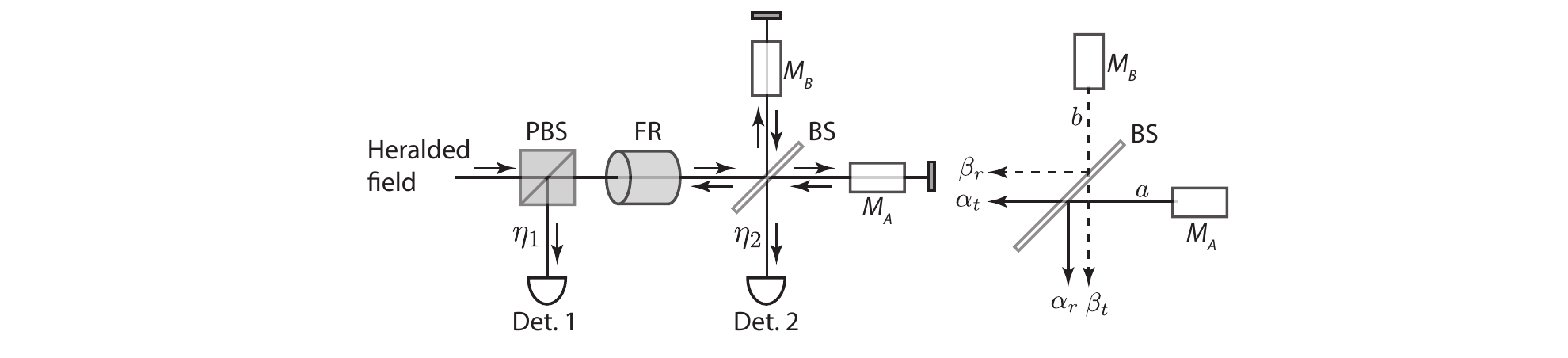}
  \caption{Experimental setup used to estimate $\bar{p}_{11}$
    directly. This setup is the same of Fig.~1 of the main text. The
    relative phase between the arms of the interferometer was
    randomized by letting the interferometer (see
    Fig.~\ref{fig:setup}) drift by itself over the total measurement
    period of more than 100~hours.  PBS: polarizing beamsplitter; FR:
    Faraday rotator; BS: beamsplitter; $M_A$ and $M_B$ are crystal
    memories~A and B. The detection efficiency from the output modes
    of the beamsplitter to detectors~1 and~2 and $\eta_1$ and
    $\eta_2$, respectively. The amplitude transmission and reflection
    coefficients of modes $a$ and $b$ right after the crystals,
    through the beamsplitter, are shown.}
  \label{fig:setup}
\end{figure*}

We now describe how
we estimate the probability $p_{11}$ from the measurement of threefold
coincidences.
For this, we use the experimental setup shown on Fig.~\ref{fig:setup};
this is the same setup shown on Fig.\@~1 of the main text. Unlike our
estimation based on the measurement of the cross-correlation
(section~\ref{section:p11gsi}), this direct method does not require
assumptions on the statistics of the source of photon pairs and on how
the memories operate.
 
Let us consider the two-photon part, denoted $\rho_2$, of the complete
density matrix describing the state of the retrieved fields just after
the crystals. If the relative phase between the spatial modes is
randomized, then the coherences are zero and we can write
\begin{equation}
  \rho_2 = q_{11} \projector{11}{11} + q_{20} \projector{20}{20} + q_{02}\projector{02}{02}. \label{eqn:rho2}
\end{equation}
Here, $q_{11}$ is the probability per heralding signal that the fields
retrieved from crystals~A and~B each contain one photon, $q_{20}$
($q_{02}$) the probability per heralding signal that the fields
retrieved from crystals~A and~B each contain 2 and 0 (0 and 2)
photons, respectively. Estimation of $p_{11}$ can be performed by
measuring the $q_{11}$ probability in Eq.~\ref{eqn:rho2}, as in
Ref.\@~\cite{Chou2005}. We show here that one can also estimate
$p_{11}$ by measuring $q_{20}$ and
$q_{02}$. 

We first derive useful relations between the occupation
probabilities. We use the following definitions: $P_H$ is the
probability to herald the presence of a signal photon (i.e.~the
probability to detect an idler photon during the detection time
window); $\eta_{echo}^A$ ($\eta_{echo}^B$) is the storage and
retrieval efficiency of crystal~A (B); $P_2$ is the probability that
the source of photon pair emits 2 pairs of photons during the
detection time window; $R$ and $T$ are the intensity reflection and
transmission coefficients of the beamsplitter when photons are
incident from the left hand side, satisfying $R+T\le 1$ and $R, T > 0$
(see Fig.\@~\ref{fig:setup}). Using these definitions, and the linear
nature of the optical storage process, we have
\begin{eqnarray*}
  q_{11} &=& \frac{1}{P_H}2 RT\eta_{echo}^A \eta_{echo}^B P_2, \\
  q_{20} &=& \frac{1}{P_H} T^2 (\eta_{echo}^A)^2 P_2, \\
  q_{20} &=& \frac{1}{P_H} R^2 (\eta_{echo}^B)^2 P_2,
\end{eqnarray*}
with which we can write $q_{20}$ and $q_{02}$ as a function of
$q_{11}$:
\begin{equation}
  \begin{array}{ccc}
    q_{20} &=& q_{11} (R/ 2T), \\
    q_{02} &=& q_{11} (T/2R).
  \end{array} \label{eqn:qs}
\end{equation}
We use these expressions later.

We now consider recombination at the beamsplitter. We describe the
effect of the beamsplitter on mode operators $a$ and $b$ as follows:
\begin{eqnarray*}
  \adag &\rightarrow& \alpha_t \adag + \alpha_r \bdag + \alpha_l \ldag, \\
  \bdag &\rightarrow& \beta_r \adag + \beta_t \bdag + \beta_l \ldag,
\end{eqnarray*}
where mode $l$ represents loss. We set all coefficients to be
real. Energy conservation yields $\alpha_r^2 + \alpha_t^2 + \alpha_l^2
= 1$ (and similarly for the coefficients of mode $b$). Also, unitarity
imposes that $\alpha_r \alpha_t + \beta_r \beta_t = 0$. Finally, we
suppose that $0<\alpha_t^2, \alpha_r^2, \beta_t^2,\beta_r^2 \le 1/2$,
and we note that $R = \beta_r^2$ and $T=\alpha_t^2$. Using these
notations, we can show that the probability to get a coincidence at
detectors~1 and~2 per heralding signal is given by
\begin{equation}
  \displaystyle \bar{p}_{11} = (a_{11} q_{11} + a_{20} q_{20} + a_{02} q_{02}) \eta_{1} \eta_{2}, \label{eqn:p}
\end{equation}
where $\eta_1$ ($\eta_2$) is the transmission from output mode $a$
($b$) -- after the beamsplitter -- to detector~1 (detector~2),
including the detector efficiency. We also have
\begin{eqnarray*}
  a_{11} &=& \beta_r^2 \left(\alpha_r -\frac{\beta_t^2}{\alpha_r}\right)^2, \\
  a_{20} &=& 2\alpha_t^2 \alpha_r^2, \\
  a_{02} &=& 2\beta_t^2 \beta_r^2. \label{eqn:as}
\end{eqnarray*}
We characterized our beamsplitter and obtained $\alpha_t^2 = T =
0.479$, $\alpha_r^2=0.422$, $\beta_t^2=0.482$ and $\beta_r^2 = R =
0.409$. These values yield
\begin{eqnarray}
  \begin{array}{ccc}
    a_{11} &=& 0.0028, \\
    a_{20} &=& 0.394, \\
    a_{02} &=& 0.404.
  \end{array} \label{eqn:as2}
\end{eqnarray}
We immediately notice that $a_{11}$ can be neglected with respect to
$a_{20}$ and $a_{02}$. This essentially means that whenever a photon
is retrieved from each crystal, they bunch with almost certainty and
do not significantly contribute to threefold coincidences. It also
requires indistinguishability of the retrieved photons, which is
consistent with the large single photon interference visibility
observed ($96.5\%$). Combining Eq.\@~\ref{eqn:p} and
Eq.\@~\ref{eqn:as}, and writing $a_{20} \approx a_{02} \approx 0.4$,
we get
\begin{equation}
  \bar{p}_{11} \approx 0.4\, \eta_{1} \eta_{2} (q_{20} + q_{02}).
\end{equation}
Finally, using Eq.\@~\ref{eqn:qs}, we get
\begin{eqnarray*}
  q_{20} + q_{02} &=& \left( \frac{R}{2T} + \frac{T}{2R} \right) q_{11} \\
  &=& 1.012\, q_{11} \\
  &\approx& q_{11}
\end{eqnarray*}
and thus
\begin{equation}
  \bar{p}_{11} \approx 0.4\, \eta_{1} \eta_{2} q_{11}. \label{eqn:preal}
\end{equation}

We now have to determine how to use Eq.\@~\ref{eqn:preal} in the
calculation of the concurrence. This method should give a result
consistent with the way we measured $p_{10} +p_{01}$, which was done
as follows. We let the phase of the interferometer drift to randomize
it, and measured the probability to get a detection at detector~1 or~2
per heralding signal. This is equivalent to, first, measure crystal~A
(with the path of crystal~B blocked) using an effective detector of
efficiency $\eta_A = \alpha_t^2 \eta_1 + \alpha_r^2 \eta_2$, and then
measure crystal~B (with the path of crystal~A blocked) using an
effective detector of efficiency $\eta_B =\beta_r^2 \eta_1 + \beta_t^2
\eta_2$, and then add the results together. Hence, if both crystals
were measured simultaneously and separately, the threefold coincidence
probability per heralding signal would be
\begin{equation}
  q_{11} \eta_A \eta_B. \label{eqn:q11ab}
\end{equation}
We can now find the relation between Eq.\@~\ref{eqn:preal}
and~\ref{eqn:q11ab}. For this, we use the fact that, in our
experiment, $\eta_2$ was larger than $\eta_1$, which is apparent from
the visibility plot in the main text (inset of Fig.\@~2-A). Hence, by
setting $\eta_1 = \eta_2/2$, we get $q_{11} \eta_A \eta_B \approx
0.454\,\eta_2^2 q_{11}$. Therefore, we immediately see that if we
multiply Eq.\@~\ref{eqn:preal} by 2.27, we get Eq.\@~\ref{eqn:q11ab}.

In summary, to estimate the probability $p_{11}$ used in the formula
of the concurrence, namely
\begin{equation}
  C \ge \max(0,V(p_{10}+p_{01})-2\sqrt{p_{00}p_{11}}), \label{eqn:C}
\end{equation}
we multiply the measured $\bar{p}_{11}$ by 2.27. This multiplicative
factor was obtained using approximations, and hence has some
uncertainty associated to it. Nevertheless, as explained in the next
subsection, this uncertainty can be neglected in front of the large
statistical uncertainty of~$\bar{p}_{11}$.

\subsection{Statistical uncertainty} \label{section:uncertainty} Here
we provide details on how the statistical uncertainty on
$\bar{p}_{11}$ was calculated in the method based on the measurement
of threefold coincidences (subsection~\ref{subsection:p11direct}).  We
denote $N_H$ the number of heralding signals during the measurement
period $T$, $\bar{P}_{11}$ the stochastic variable representing the
probability to register a threefold coincidence per heralding signal,
and $\bar{p}_{11}$ our estimation of $\bar{P}_{11}$ obtained from the
measurement.

Since $\bar{p}_{11} \ll 1$ and $N_H \gg 1$, the distribution of the
number of coincidences between detectors~1 and~2 per heralding signal
(i.e.~threefold coincidences) is accurately described by a Poisson
distribution. Specifically, the probability to get $n$ coincidences
given $N_H$ heralding signals is given by
\begin{equation}
  \label{eqn:lik}
  P(n|\bar{p}_{11})=e^{-N_H \bar{p}_{11}} \frac{(N_H
    \bar{p}_{11})^n}{n!}. 
\end{equation} 
The probability $P(n|\bar{p}_{11})$ is the likelihood function of
$\bar{p}_{11}$ given $n$. Hence, $\bar{P}_{11}$ can be estimated by
maximizing this likelihood, which yields
\begin{eqnarray*}
  \bar{p}_{11}^{\text{(MLE)}} &=& \frac{n}{N_H} \pm \frac{\sqrt{n}}{N_H},
\end{eqnarray*}
where MLE stands for ``maximum likelihood estimation''. This method
yields an appropriate estimate of $\bar{P}_{11}$, unless one measures
$n=0$, in which case one would conclude that $\bar{p}_{11} = 0\pm 0$.

A more conservative estimate can be obtained in the following
way. Using Bayes' law, we can write the probability density function
$\Omega (\bar{p}_{11}|n)$ of getting $\bar{P}_{11} = \bar{p}_{11}$
given $n$ measured threefold coincidences:
$$\Omega(\bar{p}_{11}|n)=\frac{P(n|\bar{p}_{11})}{\int_0^1 P(n|\bar{p}_{11})\, d\bar{p}_{11}}.$$
The denominator of Eq.~\ref{eqn:lik} can be calculated:
\begin{eqnarray*}
  \int_0^1 e^{-N_H \bar{p}_{11}} \frac{(N_H \bar{p}_{11})^n}{n!}\, d\bar{p}_{11} &=& \frac{1}{N_H}\left( 1-e^{-N_H} \sum_{k=0}^{n}\frac{N_H^k}{k!}
  \right) \\
  &\approx& \frac{1}{N_H}.
\end{eqnarray*}
The last equality holds because $n \ll N_H$.  Therefore:
\begin{equation}
  \Omega(\bar{p}_{11}|n)= N_H P(n|\bar{p}_{11}). \label{eqn:omega}
\end{equation}

Using Eq.~\ref{eqn:omega}, we can calculate the expected value of
$\bar{P}_{11}$ and its second moment, respectively denoted $\langle
\bar{P}_{11} \rangle$ and~$\langle \bar{P}_{11}^2 \rangle$:
\begin{eqnarray*}
  \langle \bar{P}_{11} \rangle &=& \int_0^1 \bar{p}_{11} \, \Omega(\bar{p}_{11}|n) \,d\bar{p}_{11} \approx \frac{n+1}{N_H},\\
  \langle \bar{P}_{11}^2 \rangle &=& \int_0^1 \bar{p}_{11}^2 \, \Omega(\bar{p}_{11}|n) \,d\bar{p}_{11} \approx \frac{(n+1)(n+2)}{N_H^2}. \\
\end{eqnarray*}

\noindent The standard deviation of $\bar{P}_{11}$ is therefore
\begin{eqnarray*}
  \sqrt{\langle \bar{P}_{11}^2 \rangle - \langle \bar{P}_{11} \rangle^2} = \frac{\sqrt{n+1}}{N_H}.
\end{eqnarray*}

Hence, given $n$ measured threefold coincidences and $N_H$ heralding
signals, this conservative method yields
\begin{equation*}
  \bar{p}_{11}^{\text{(CE)}}= \frac{n+1}{N_H} \pm \frac{\sqrt{n+1}}{N_H},
\end{equation*}
where CE stands for ``conservative estimation''. In our results, we
estimate the concurrence with both the maximum likelihood and the
conservative estimations of $\bar{P}_{11}$.

\subsection{Experimental results} \label{subsection:results}

The estimation of $p_{00}$, $p_{10}+p_{01}$ and $p_{11}$ proceeded as
follows. For a time $T=166$~hours, we let the interferometer drift to
randomize the phase. During that period of time, we recorded the total
number of heralded twofold coincidences between the heralding detector
and detector~1 ($N_{1|H}$), and between the heralding detector and
detector~2 ($N_{2|H}$), and calculated $N_2 = N_{1|H} + N_{2|H}$. We
also recorded the total number of heralded threefold coincidences,
denoted $N_{12|H}$, between all three detectors. We used a coincidence
time window of 10~ns; this is the same value we used in the estimation
of the concurrence based on the cross-correlation. We calculated
$$p_{10}+p_{01} = \frac{N_2}{N_H}\pm\frac{\sqrt{N_2}}{N_H}.$$ 
For $p_{11}$, we used both the MLE, yielding
\begin{eqnarray*}
  p_{11}^{(\text{MLE})} = 2.27 \left( \frac{N_{12|H}}{N_H} \pm \frac{\sqrt{N_{12|H}}}{N_H} \right),
\end{eqnarray*}
and the CE:
\begin{equation*}
  p_{11}^{(\text{CE})} = 2.27 \left( \frac{N_{12|H}+1}{N_H} \pm \frac{\sqrt{N_{12|H}+1}}{N_H} \right).
\end{equation*}
Finally, we estimated $p_{00} = 1- p_{10}-p_{01} - p_{11}$.

With a 10~ns coincidence window, we obtained $p_{10}+p_{01} =
1.7777(34) \times 10^{-4}$, $V=96.5(1.2)\%$, and observed two
threefold coincidences ($N_{12|H}=2$). This yields
\begin{eqnarray*}
  p_{11}^{(\text{MLE})} &=& 2.9(2.1)\times 10^{-9} \\
  C^{(\text{MLE})} &\ge& 6.3(3.8) \times 10^{-5}
\end{eqnarray*}
and
\begin{eqnarray*}
  p_{11}^{(\text{CE})} &=& 3.9(2.2)\times 10^{-9} \\
  C^{(\text{CE})} &\ge& 3.9(3.8) \times 10^{-5}
\end{eqnarray*}
Both the MLE and CE yield a lower bound for the concurrence that is
greater than~0 by at least one standard deviation.

\bibliography{heralded-nomonths}

\end{document}